\documentclass[aps,letterpaper,twocolumn,prl,showpacs,nofootinbib]{revtex4-1}

\usepackage{amssymb}
\usepackage{amsmath,bm}

\usepackage[pdftex]{graphicx,color}
\usepackage[english]{babel}
\usepackage{hyperref}
\usepackage{enumitem}
\usepackage[dvipsnames]{xcolor}
\usepackage{epstopdf}
\usepackage{tikz}

\begin{document}

\title{Parametrically driven pure-Kerr temporal solitons in a chip-integrated microcavity}

\author{Gr\'egory Moille$^{1,2}$}
\author{Miriam Leonhardt$^{3,4}$}
\author{David Paligora$^{3,4}$}
\author{Nicolas Englebert$^{5}$}
\author{Fran\c{c}ois Leo$^{5}$}
\author{Julien Fatome$^{6}$}
\author{Kartik Srinivasan$^{1,2}$}
\author{Miro Erkintalo$^{3,4}$}
\email{m.erkintalo@auckland.ac.nz}

\affiliation{$^1$Joint Quantum Institute, NIST/University of Maryland, College Park, USA}
\affiliation{$^2$Microsystems and Nanotechnology Division, National Institute of Standards and Technology, Gaithersburg, USA}
\affiliation{$^3$Department of Physics, University of Auckland, Auckland 1010, New Zealand}
\affiliation{$^4$The Dodd-Walls Centre for Photonic and Quantum Technologies, New Zealand}
\affiliation{$^5$Service OPERA-Photonique, Universit\'e libre de Bruxelles (U.L.B.), 50 Avenue F. D. Roosevelt, CP 194/5, B-1050 Brussels, Belgium}
\affiliation{$^6$Laboratoire Interdisciplinaire Carnot de Bourgogne, UMR 6303 CNRS Universit\'e de Bourgogne, Dijon, France}

\begin{abstract}
\noindent The discovery that externally-driven nonlinear optical resonators can sustain ultrashort pulses corresponding to coherent optical frequency combs has enabled landmark advances in applications from telecommunications to sensing. The main research focus has hitherto been on resonators with purely cubic (Kerr-type) nonlinearity that are externally-driven with a monochromatic continuous wave laser -- in such systems, the solitons manifest themselves as unique attractors whose carrier frequency coincides with that of the external driving field. Recent experiments have, however, shown that a qualitatively different type of temporal soliton can arise via parametric down-conversion in resonators with simultaneous quadratic and cubic nonlinearity. In contrast to conventional solitons in pure-Kerr resonators, these \emph{parametrically driven solitons} come in two different flavours with opposite phases, and they are spectrally centred at half of the frequency of the driving field. Here, we theoretically predict and experimentally demonstrate that parametrically driven solitons can also arise in resonators with pure Kerr nonlinearity under conditions of bichromatic driving. In this case, the solitons arise through four-wave mixing mediated phase-sensitive amplification, come with two distinct phases, and have a carrier frequency in between the two external driving fields. Our experiments are performed in an integrated silicon nitride microcavity, and we observe frequency comb spectra in good agreement with theoretical predictions. In addition to representing a fundamental discovery of a new type of temporal dissipative soliton, our results constitute the first unequivocal realisation of parametrically driven soliton frequency combs in a microcavity platform compatible with foundry-ready mass fabrication.
\end{abstract}

\maketitle

\section{Introduction}
\begin{figure*}[!t]
	\centering
	\includegraphics[width = \textwidth, clip=true]{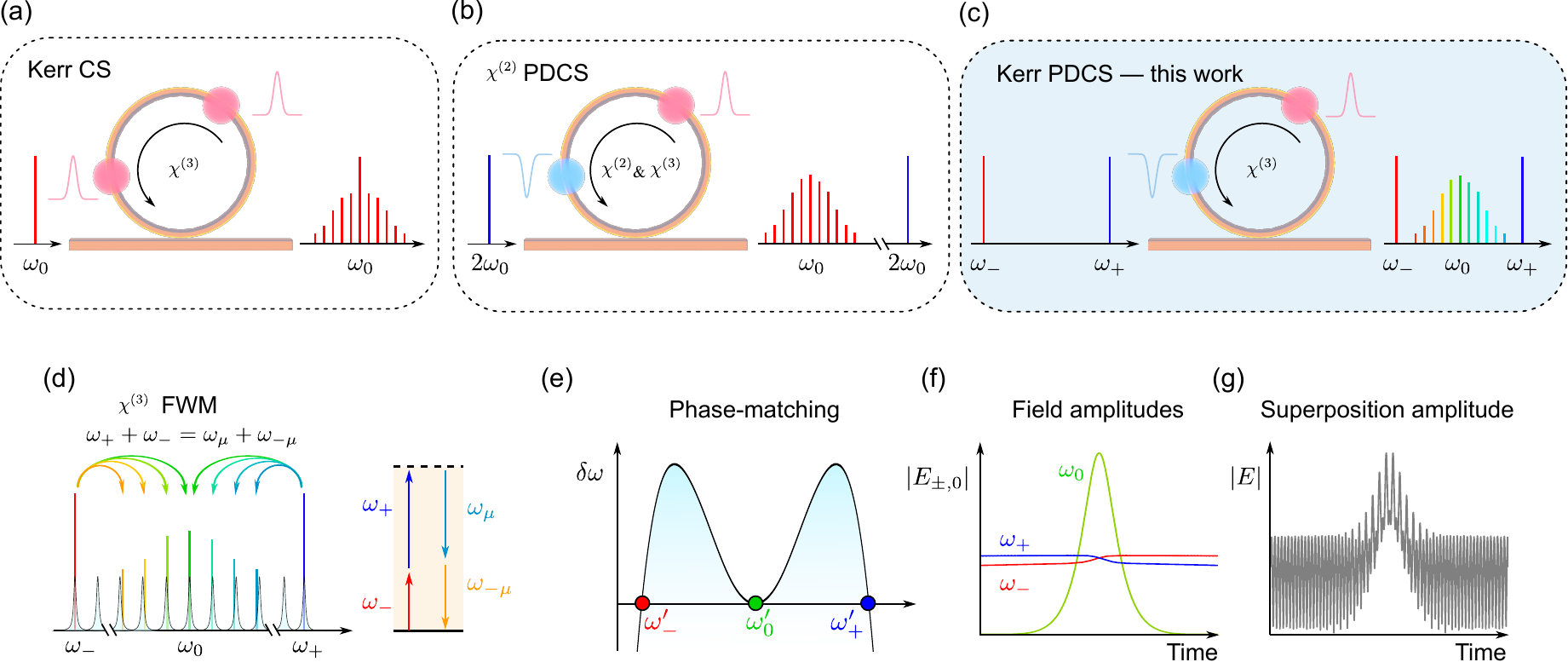}
	\caption{\textbf{Comparison of platforms and schematic illustration of PDCS generation in Kerr resonators.} %
	(a) Conventional Kerr CSs~\cite{leo_temporal_2010,herr_temporal_2014,kippenberg_dissipative_2018} form around the input frequency $\omega_0$ in dispersive resonators with $\chi^{(3)}$ Kerr nonlinearity; for given parameters, all solitons in the resonator are identical. (b) Parametric down-conversion of an input field at $2\omega_0$ can yield PDCSs at $\omega_0$ in a resonator with combined $\chi^{(2)}$ and $\chi^{(3)}$ nonlinearity~\cite{englebert_parametrically_2021, bruch_pockels_2021}. Here, the solitons come in two forms with opposite phase~\cite{englebert_parametrically_2021}: the complex soliton electric field $E_\pm(\tau)\propto \pm E_0(\tau)\exp(i\omega_0 \tau)$, where $\pm E_0(\tau)$ is the slowly-varying envelope [real part visualised in (a)--(c)]. (c) In the bichromatically driven Kerr resonator configuration studied in this work, PDCSs arise in between two input frequencies $\omega_\pm$. (d) Cartoon of the non-degenerate FWM process (and corresponding energy flow diagram) through which the intracavity fields at $\omega_\pm$ provide coherent parametric driving to \emph{all} of the PDCS comb lines around $\omega_0$. Note that this energy flow is in contrast to the standard Kerr CS case in (a), where only the mode $\omega_0$ is driven. The shaded curves in the background of (d) depict cavity modes. (e) PDCSs arise under conditions close to linear phase-matching of degenerate FWM, which in terms of cavity modes occurs when the frequency deviation $\delta\omega = (\omega'_+ + \omega'_-) - \omega'_0\approx 0$, with $\omega'_\pm$ the driven cavity modes and $\omega'_0$ the mode closest to $\omega_0$. (f) Illustrative slowly-varying electric field amplitudes around the parametric signal frequency $\omega_0$ ($E_0$, green) and the pump frequencies $\omega_\pm$ ($E_\pm$, blue and red). The fields $E_\pm$ must be approximately CW to ensure homogeneous parametric driving strength, calling for (i) sufficient dispersive walk-off to mitigate pump depletion and (ii) suppression of modulation instabilities. (g) Because the full intracavity amplitude consists of a superposition of the $E_\pm$ and $E_0$ fields, the PDCS manifests itself as a localized structure amidst a rapidly oscillating background.}
	\label{fig1}
\end{figure*}

The injection of monochromatic continuous wave (CW) laser light into dispersive optical resonators with purely Kerr-type $\chi^{(3)}$ nonlinearity can lead to the generation of localized structures known as dissipative Kerr cavity solitons (CSs)~\cite{leo_temporal_2010, kippenberg_dissipative_2018}. These CSs correspond to ultrashort pulses of light that can persist within the resonator [Fig.~\ref{fig1}(a)], indefinitely maintaining constant shape and energy~\cite{wabnitz_suppression_1993}. While first observed in macroscopic optical fiber ring resonators~\cite{leo_temporal_2010}, CSs have attracted particular attention in the context of monolithic Kerr microcavities~\cite{kippenberg_dissipative_2018}, where they underpin the generation of coherent and broadband optical frequency combs~\cite{herr_temporal_2014,brasch_photonic_2016,pasquazi_micro-combs_2018}. By offering a route to coherent frequency comb generation in chip-integrated, foundry-ready platforms, CSs have enabled ground breaking advances in applications including telecommunications~\cite{marin-palomo_microresonator-based_2017, corcoran_ultra-dense_2020}, artificial intelligence~\cite{xu_11_2021, feldmann_parallel_2021}, astronomy~\cite{suh_searching_2019, obrzud_microphotonic_2019}, frequency synthesis~\cite{spencer_optical-frequency_2018}, microwave generation~\cite{lucas_ultralow-noise_2020, kwon_ultrastable_2022}, and distance measurements~\cite{suh_soliton_2018, riemensberger_massively_2020}.

The conventional CSs that manifest themselves in resonators with pure Kerr nonlinearity sit atop a CW background, and they gain their energy through four-wave-mixing (FWM) interactions with that background~\cite{leo_temporal_2010}. In the frequency domain, the solitons are (to first order) centred around the frequency of the external CW laser that drives the resonator [Fig.~\ref{fig1}(a)]. They are (barring some special exceptions~\cite{nielsen_coexistence_2019,anderson_coexistence_2017,hansson_frequency_2015,xu_spontaneous_2021,lucas_spatial_2018}) unique attracting states: except for trivial time translations, all the CSs that exist for given system parameters are identical. These features can be disadvantageous or altogether prohibitive for selected applications: noise on the external CW laser can degrade the coherence of nearby comb lines, removal of the CW background may require careful spectral filtering,  whilst applications that require coexistence of distinguishable binary elements~\cite{takesue_10_2016,okawachi_quantum_2016,okawachi_dynamic_2021, inagaki_Large-scale_2016, mohseni_ising_2022} are fundamentally beyond reach. Interestingly, recent experiments reveal that qualitatively different types of CSs can exist in resonators that display a quadratic $\chi^{(2)}$ in addition to a cubic $\chi^{(3)}$ nonlinearity [Fig.~\ref{fig1}(b)]; in particular, degenerate optical parametric oscillators driven at $2\omega_0$ can support CSs at $\omega_0$~\cite{englebert_parametrically_2021, bruch_pockels_2021}. In this configuration, the solitons are \emph{parametrically driven} through the quadratic down-conversion of the externally-injected field, which endows them with fundamental differences compared to the conventional CSs emerging in monochromatically-driven, pure-Kerr resonators. Specifically, \emph{parametrically driven cavity solitons} (PDCSs) are spectrally separated from the driving frequency (e.g. $\omega_0$ versus $2\omega_0$), and they come in two binary forms with opposite phase. These traits render PDCSs of interest for an altogether new range of applications.

Optical PDCSs have so far been generated only via the quadratic $\chi^{(2)}$ nonlinearity, which is not intrinsically available in integrated (foundry-ready) resonator platforms, such as silicon~\cite{thomson_roadmap_2016} or silicon nitride~\cite{liu_high-yield_2021,luSHG2020, NitissNat.Photon.2022}. However, it is well-known that phase-sensitive amplification analogous to $\chi^{(2)}$ parametric down-conversion can also be realised in pure Kerr resonators when driven with two lasers with different carrier frequencies~\cite{mecozzi_long-term_1994,agrawal_nonlinear_nodate,radic_two-pump_2003, okawachi_dual-pumped_2015, Andrekson_fiber-based_2020}, allowing e.g. for novel random number generators~\cite{takesue_10_2016,okawachi_quantum_2016,okawachi_dynamic_2021} and coherent optical Ising machines~\cite{inagaki_Large-scale_2016, mohseni_ising_2022}. A natural question that arises is: is it possible to generate PDCSs in foundry-ready, pure-Kerr resonators with bichromatic driving? Whilst a related question has been theoretically explored in the context of \emph{diffractive} Kerr-only resonators~\cite{de_valcarcel_phase-bistable_2013}, the presence of \emph{dispersion} substantially changes the physics of the problem. The impact of bichromatic driving in the dynamics of conventional Kerr CSs has also been considered~\cite{hansson_bichromatically_2014, ceoldo_multiple_2016, zhang_spectral_2020,moille_ultra-broadband_2021, qureshi_soliton_2021, taheri_all-optical_2022}, but the possibility of using the scheme to generate temporal PDCSs remains unexplored.

Here, we theoretically predict and experimentally demonstrate that a dispersive resonator with pure Kerr nonlinearity can support PDCSs in the presence of bichromatic driving [Fig.~\ref{fig1}(c)]. We reveal that, under appropriate conditions, a signal field with carrier frequency in between two spectrally-separated driving fields obeys the damped, parametrically driven nonlinear Schr\"odinger equation (PDNLSE) that admits PDCS solutions, and we unveil the system requirements for the practical excitation of such solutions. Our experiments are performed in a 23~$\mu$m-radius, chip-integrated silicon nitride microring resonator whose dispersion is judiciously engineered to facilitate PDCS generation at 253~THz (1185~nm) when bichromatically pumping at 314~THz (955~nm) and 192~THz (1560~nm). We observe PDCS frequency comb spectra that are in good agreement with numerical simulations, as well as clear signatures of the anticipated $\mathbb{Z}_2$ symmetry, i.e., coexistence of two PDCSs with opposite phase. By revealing a fundamentally new pathway for the generation of coherent PDCS frequency combs far from any pump frequency, in a platform that has direct compatibility with foundry-ready fabrication, our work paves the way for integrated, low-noise frequency comb generation in new spectral regions, as well as photonic integration of applications requiring combs with a binary degree of freedom.

\section{Results}
We first summarise the main points that lead to the prediction of PDCSs in bichromatically-driven Kerr resonators [for full details, see Methods]. To this end, we consider a resonator made out of a dispersive, $\chi^{(3)}$ nonlinear waveguide that is driven with two coherent CW fields with angular frequencies $\omega_\pm$ [see Fig.~\ref{fig1}(c)]. The dispersion of the resonator is described by the integrated dispersion~\cite{pasquazi_micro-combs_2018} at the cavity resonance $\omega'_0$ (apostrophes highlight resonance frequencies throughout the article) closest to the frequency \mbox{$\omega_0 = (\omega_+ + \omega_-)/2$}:
\begin{equation}
D_\mathrm{int}(\mu) = \omega'_\mu - \omega'_0 - \mu D_1 = \sum_{k\geq 2} \frac{D_k}{k!} \mu^k. \label{Dint}
\end{equation}
Here, $\mu$ is a relative mode number with respect to the resonance $\omega_0'$ and $D_1/(2\pi)$ is the cavity free-spectral range (FSR) at $\omega'_0$. The terms $D_k$ with $k>1$ account for deviations of the resonance frequencies $\omega'_\mu$ from an equidistant grid defined by $\omega'_0 + \mu D_1$.

Under particular conditions [see Methods], the evolution of the slowly-varying electric field envelope centred at $\omega_0$ can be shown to be (approximately) governed by the PDNLSE, with the parametric driving ensuing from non-degenerate FWM driven by the intracavity fields at the pump frequencies [$\omega_+ + \omega_- \rightarrow \omega_\mu + \omega_{-\mu}$, see Fig.~\ref{fig1}(d)]. (Note: in stark contrast to standard Kerr CSs, for which only \emph{one} comb line is externally driven, \emph{all} of the components of a PDCS frequency comb are separately driven via non-degenerate FWM.) Because the PDNLSE is well-known to admit PDCS solutions~\cite{englebert_parametrically_2021, miles_parametrically_1984, barashenkov_stability_1991}, it follows that the system may support such solitons with a carrier frequency $\omega_0$ in between the two driving frequencies, provided however that the system parameters -- particularly resonator dispersion -- are conducive for soliton existence.

\begin{figure*}[!t]
	\centering
	\includegraphics[width = \textwidth, clip=true]{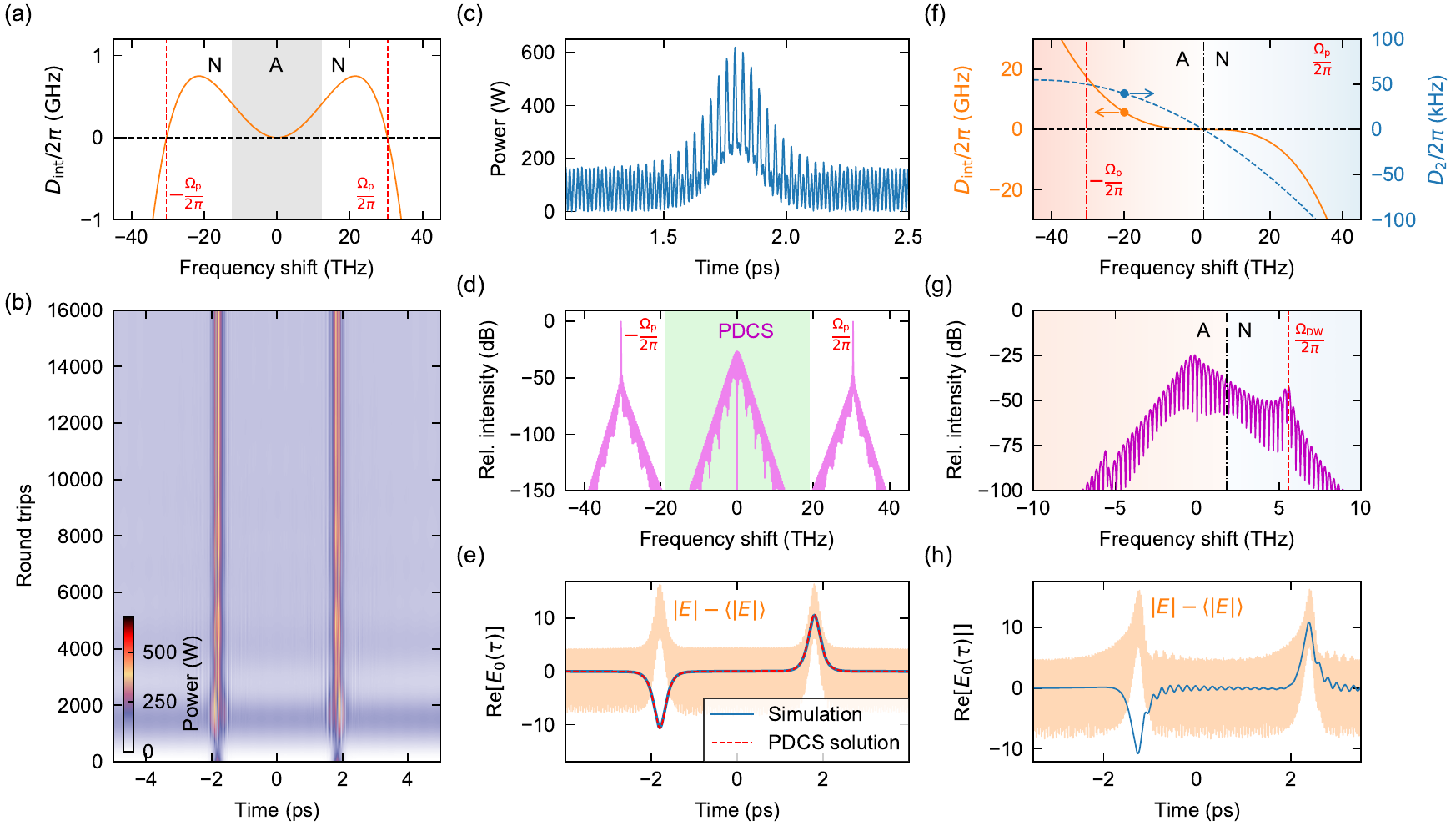}
	\caption{\textbf{Illustrative simulations of PDCSs in dispersive Kerr resonators.} (a)--(e) Simulation results obtained for a 25~GHz toy resonator with $D_2 = 2\pi\times 4.1~\mathrm{kHz}$ and $D_4 = -2\pi\times 33~\mathrm{mHz}$, yielding the integrated dispersion in (a). The shaded gray region highlights a region of anomalous (A) dispersion sandwiched between regions of normal (N) dispersion. Note that, because $D_\mathrm{int}$ is completely symmetric in this example, the frequency mismatch $\delta\omega(p) = D_\mathrm{int}(p)$, such that the pump frequencies satisfying linear phase-matching $\delta\omega \approx 0$ can be directly read off the graph: the dashed vertical lines in (a) indicate those pump frequencies and were used in the simulations. (b) Dynamical evolution of two hyperbolic secant pulses with opposite phase. The colormap depicts instantaneous power in Watts. (c) Temporal intensity profile around one of the steady-state solitons at the output of the simulation in (b). (d) Optical spectrum corresponding to the output of the simulation in (b). (e) The red dashed curve shows the real part of the analytical PDCS solution [see Methods] while the solid blue curve shows the real part of the simulated intracavity field about zero frequency shift. The simulation result was obtained by first spectrally filtering out the intracavity fields at the pump frequencies (green dashed area in (d) indicates the filter passband). The orange curve shows the (mean-subtracted) total field amplitude for reference. (f)--(h) Simulation results with parameters as in (a)--(e) but with an additional third-order dispersion term $D_3 = -2\pi\times 58~\mathrm{Hz}$, yielding the integrated dispersion (left axis) and corresponding group-velocity dispersion $D_2$ (right axis) shown in (f). (g) PDCS spectrum in the presence of third-order dispersion; vertical red dashed line indicates the predicted dispersive wave position. The black vertical dash-dotted line in (f) and (g) indicate the zero-dispersion point that demarcates regions of normal (N) and anomalous (A) dispersion. (h) The blue and orange curves are as in (e) but with third-order dispersion. No analytical solution exist in the presence of third-order dispersion.}
	\label{fig2}
\end{figure*}

The resonator dispersion must meet three key conditions for PDCS excitation to be viable [Methods]. First, for solitons to exist, the dispersion around the degenerate FWM frequency $\omega_0$ must be anomalous, i.e., $D_2 > 0$ in Eq.~\eqref{Dint}. Second, the effective detuning [see Methods] between the degenerate FWM frequency ($\omega_0$) and the closest cavity resonance ($\omega'_0$) must be within the range of soliton existence, essentially requiring that the degenerate FWM process $\omega_+ + \omega_- \rightarrow 2\omega_0$ (approximately) satisfies linear phase-matching [Fig.~\ref{fig1}(e)]. This second condition can be written as \mbox{$\delta\omega = (\omega'_++\omega'_-)/2-\omega_0' = [D_\mathrm{int}(p) + D_\mathrm{int}(-p)]/2\approx 0$}, where $\pm p$ correspond to the modes excited by the driving lasers at $\omega_\pm$. Given that $D_2 > 0$, this requires at least one higher-even-order dispersion coefficient (e.g. $D_4$) to be negative. Third, the intracavity field amplitudes at the driving frequencies, $|E_\pm|$, must remain (approximately) homogeneous and stationary to ensure a constant parametric driving strength for the PDCS field $E_0$ centred at $\omega_0$ [Fig.~\ref{fig1}(f)]. This final condition can be met by ensuring dispersion at the driving frequencies is (i) normal (or driving amplitudes small), such that the corresponding intracavity fields do not undergo pattern forming (modulation) instabilities~\cite{coen_universal_2013}, and (ii) such that the temporal walk-off between the driving frequencies $\omega_\pm$ and the signal frequency $\omega_0$ is sufficiently large so as to mitigate pump depletion in the vicinity of the soliton that would otherwise break the homogeneity of the fields at $\omega_\pm$ [Fig.~\ref{fig1}(f)].  As will be demonstrated below, all of these conditions can be met through judicious dispersion engineering that is within the reach of contemporary microphotonic fabrication.

\vskip 5pt

\noindent \textbf{Simulations.} Before discussing our experiments, we present results from numerical simulations that illustrate the salient physics. Our simulations are based upon a full iterative ``Ikeda'' map of the system without any approximations [Methods], and they consider a toy resonator with 25~GHz FSR and minimal dispersion necessary for PDCS existence [see Fig.~\ref{fig2}(a)]. Specifically, we assume a quartic dispersion with $D_2 = 2\pi\times 4.1~\mathrm{kHz}$ and $D_4 = -2\pi\times 33~\mathrm{mHz}$, yielding $D_\mathrm{int}(p) + D_\mathrm{int}(-p)\approx 0$ for pump frequency shift $\Omega_\mathrm{p} = 2\pi\times 30.4~\mathrm{THz}$ (corresponding to mode number $p = 1217$). We assume for simplicity that the two driving fields are coincident on their respective linear cavity resonances (zero detuning), and both carry CW laser power of about 140~mW [see Methods for other parameters]. Because the group-velocity dispersion at the pump frequencies is \emph{normal}, modulational instabilities are suppressed and the intracavity fields converge to stable homogeneous states with equal circulating CW power of about 43~W, thus yielding an effective parametric driving strength and detuning within the regime of PDCS existence [see Methods].

Figure~\ref{fig2}(b) shows the evolution of the numerically simulated intracavity intensity profile with an initial condition consisting of two hyperbolic secant pulses with opposite phases. As can be seen, after a short transient, the field reaches a steady-state that is indicative of two pulses circulating around the resonator. The pulses sit atop a rapidly oscillating background that is due to the beating between the quasi-homogeneous fields at the pump frequencies [Fig.~\ref{fig2}(c)]. Correspondingly, the spectrum of the simulation output [Fig.~\ref{fig2}(d)] shows clearly the presence of a hyperbolic secant-shaped feature that sits in between the strong quasi-monochromatic components at the pump frequencies. In accordance with PDCS theory [see Methods], there is no significant CW peak at the parametric signal frequency $\omega_0$ at which the solitons are spectrally centred. To highlight the phase disparity of the steady-state pulses, we apply a numerical filter to remove the quasi-monochromatic intracavity components around the pump frequencies, and plot in Fig.~\ref{fig2}(e) the real part of the complex intracavity electric field envelope. The simulation results in Fig.~\ref{fig2}(e) are compared against the real parts of the exact, analytical PDCS solutions [Methods], and we clearly observe excellent agreement.

The results in Fig.~\ref{fig2}(a)--(e) corroborate the fundamental viability of our scheme. However, they were obtained assuming a completely symmetric dispersion profile with  no odd-order terms, which may be difficult to realise even with state-of-the-art microphotonic fabrication (including the resonators considered in our experiments). We find, however, that PDCSs can exist even in the presence of odd-order-dispersion, albeit in a perturbed form. This point is highlighted in Figs.~\ref{fig2}(f)--(h), which show results from simulations with all parameters as in Fig.~\ref{fig2}(a)--(e) except an additional non-zero third-order dispersion term $D_3 = -2\pi\times 58~\mathrm{Hz}$. As for conventional (externally-driven) Kerr CSs~\cite{coen_modeling_2013, jang_observation_2014,milian_soliton_2014, brasch_photonic_2016}, we find that third-order dispersion causes the solitons to emit dispersive radiation at a spectral position determined by the phase-matching condition $D_\mathrm{int}(\mu_\mathrm{DW})\approx (\omega_0-\omega_0')$ [Fig.~\ref{fig2}(g)]. This emission results in the solitons experiencing constant drift in the temporal domain, and endows them with oscillatory tails [Fig.~\ref{fig2}(h)]. Yet, as can clearly be seen, the PDCSs continue to exist in two distinct forms with near-opposite phase. It is worth noting that, for the parameters considered in Fig.~\ref{fig2}(f)--(h), the low-frequency driving field experiences anomalous group-velocity dispersion; however, the intracavity intensity at that frequency is below the modulation instability threshold~\cite{coen_universal_2013}, thus allowing the corresponding field to remain quasi-homogeneous (the modulation on the total intensity profile arises solely from the linear beating between the different fields).\\

\begin{figure*}[!t]
	\centering
	\includegraphics[width = \textwidth, clip=true]{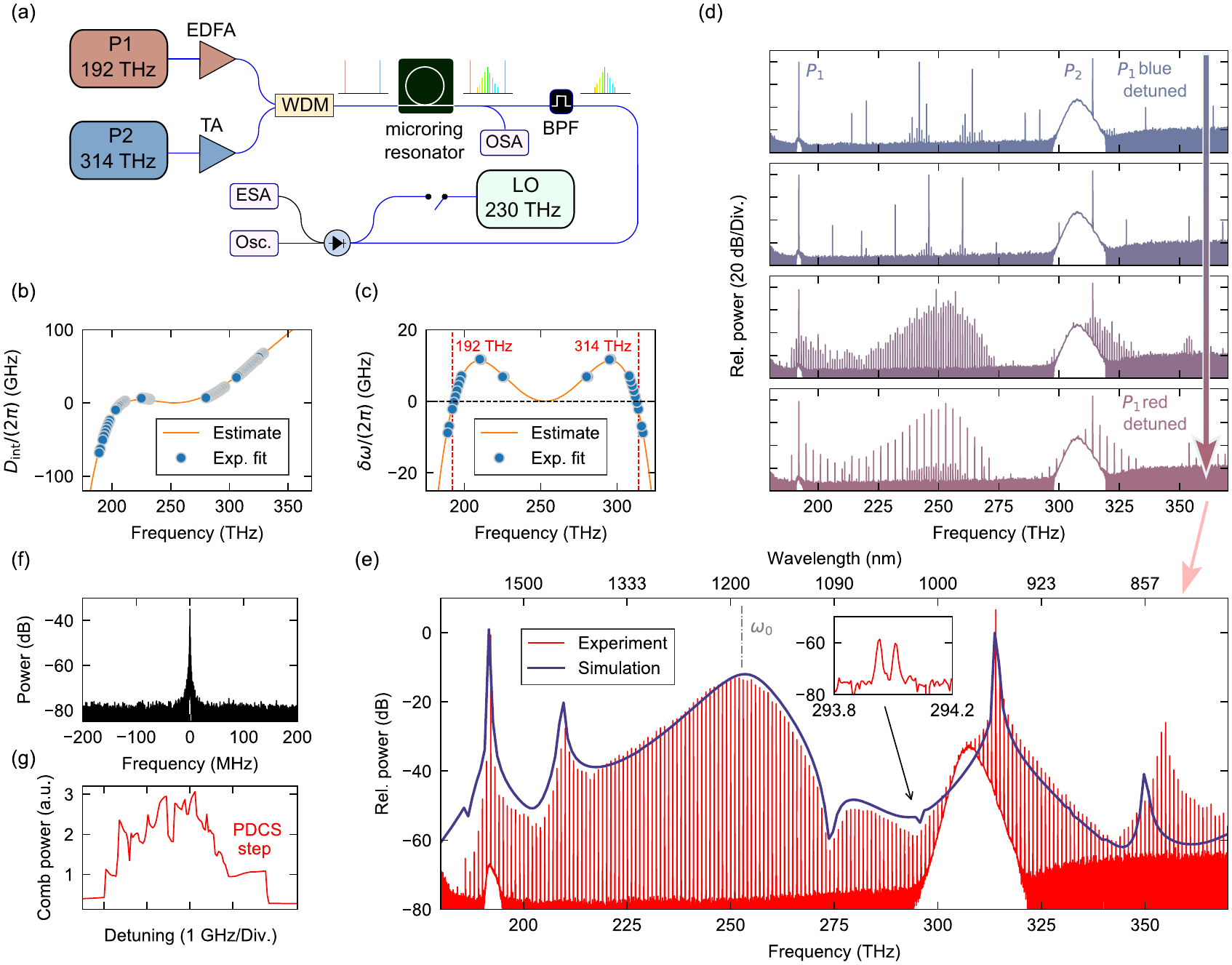}
	\caption{\textbf{Experimental observation of pure-Kerr temporal PDCSs in an on-chip microcavity.} (a) Experimental setup. EDFA, erbium-doped fiber amplifier; TA, semiconductor taper amplifier; WDM, wavelength division multiplexer; OSA, optical spectrum analyzer; BPF, band-pass filter; LO, local oscillator; ESA, electrical spectrum analyzer; Osc., oscilloscope. (b) Orange curve shows integrated dispersion around 253~THz used to simulate our experiments while blue circles show the dispersion fitted from experimental data [see Methods]. (c) Linear phase-mismatch for the degenerate FWM process computed from the integrated dispersion data in (b). The fit uncertainties for (b) and (c) are smaller than the circle markers shown [Methods]. (d) Experimentally measured spectra as the low-frequency pump (P1) tunes into resonance from the blue (high-frequency pump P2 kept fixed). (e) As the low-frequency pump tunes in sufficiently, the frequency comb spectrum abruptly transitions into a smooth envelope. This transition is indicative of a PDCS comb; the experimentally measured comb spectrum is in good agreement with a numerically simulated PDCS spectrum (blue curve, see Methods). The inset in (e) highlights the offset between the PDCS frequency comb and the comb around the P2 pump frequency. The arrows across (d) and (e) highlight the red-shift of the pump P1. (f) Heterodyne beat note observed in the PDCS regime (instrument resolution bandwidth is 10~kHz). (g) Photodetector signal as the low-frequency pump is tuned across a resonance, revealing a step feature that coincides with the emergence of the smooth PDCS comb envelope in (e). }
	\label{fig3}
\end{figure*}

\noindent\textbf{Experiments.} For experimental demonstration [see Fig.~\ref{fig3}(a) and Methods], we use a microring resonator made from a 690~nm-thick, 850~nm-wide silicon nitride layer embedded in fused silica, fabricated in a commercial foundry. The ring exhibits a radius of 23~$\mu$m, thus yielding a free-spectral range of about 1~THz. We use two external cavity diode lasers to drive the resonator: one tunable in the telecommunications C-band (from 186~THz to 198~THz, i.e., from 1613~nm to 1515~nm) and the other tunable from 306~THz to 330~THz (980~nm to 910~nm). Both driving fields are optically amplified and combined using a wavelength-division multiplexer (WDM) before being coupled into the resonator via a pulley scheme that ensures efficient coupling at all the relevant frequencies~\cite{moille_broadband_2019}. At the output of the resonator, 90\% of the signal is routed to an optical spectrum analyzer for analysis. The remaining 10\% is passed through a bandpass filter to remove spectral components around the driving frequencies, thus allowing to isolate the parametrically-generated signal field for characterisation.

The orange curve in Fig.~\ref{fig3}(b) depicts an estimate of the resonator's integrated dispersion around a cavity mode at 253~THz, obtained through a combination of finite-element-modelling and fitting to our experimental observations [see Methods]. This data is \emph{consistent} with experimentally measured resonance frequencies (blue circles), yet we caution that our inability to probe the resonances around 253~THz prevents unequivocal evaluation of the dispersion at that frequency. The estimated dispersion can be seen to be such that the requisite phase-matching for generating a PDCS at 253~THz ($\delta\omega\approx 0$) can be satisfied, provided that the pump lasers are configured to drive cavity modes at 314~THz and 192~THz [Fig.~\ref{fig3}(c)].

In our experiments, we set the on-chip driving power for both driving fields to be about 150~mW and tune the high-frequency pump to the cavity mode at 314~THz. We then progressively tune the low-frequency pump to the cavity mode at 192~THz (from blue to red), maintaining the high-frequency pump at a fixed frequency. As the low-frequency pump tunes into resonance, we initially observe non-degenerate parametric oscillation characterised by the generation of two CW components symmetrically detuned about 253~THz. These CW components progressively shift closer to each other as the pump tunes into the resonance, concomitant  with the formation of a frequency comb around the degenerate FWM frequency $\omega_0$ [see Fig.~\ref{fig3}(c)]. To characterise the comb noise, we performed a heterodyne beat measurement using a helper laser at 230~THz within the vicinity of a single comb line. Initially, no beat note is observed, which is characteristic of an unstable, non-solitonic state within the resonator. Remarkably, as the 192~THz driving field is tuned further into resonance, we observe that the parametric signals reach degeneracy, concomitant with the emergence of a broadband comb state with smooth spectral envelope [Fig.~\ref{fig3}(e)] and a heterodyne beat note (comparable with the helper laser linewidth of 250~kHz) that is considerably narrower than the 300~MHz microcavity linewidth [Fig.~\ref{fig3}(f) and Supplementary Figure~1].

\begin{figure}[!t]
	\centering
	\includegraphics[width = \columnwidth, clip=true]{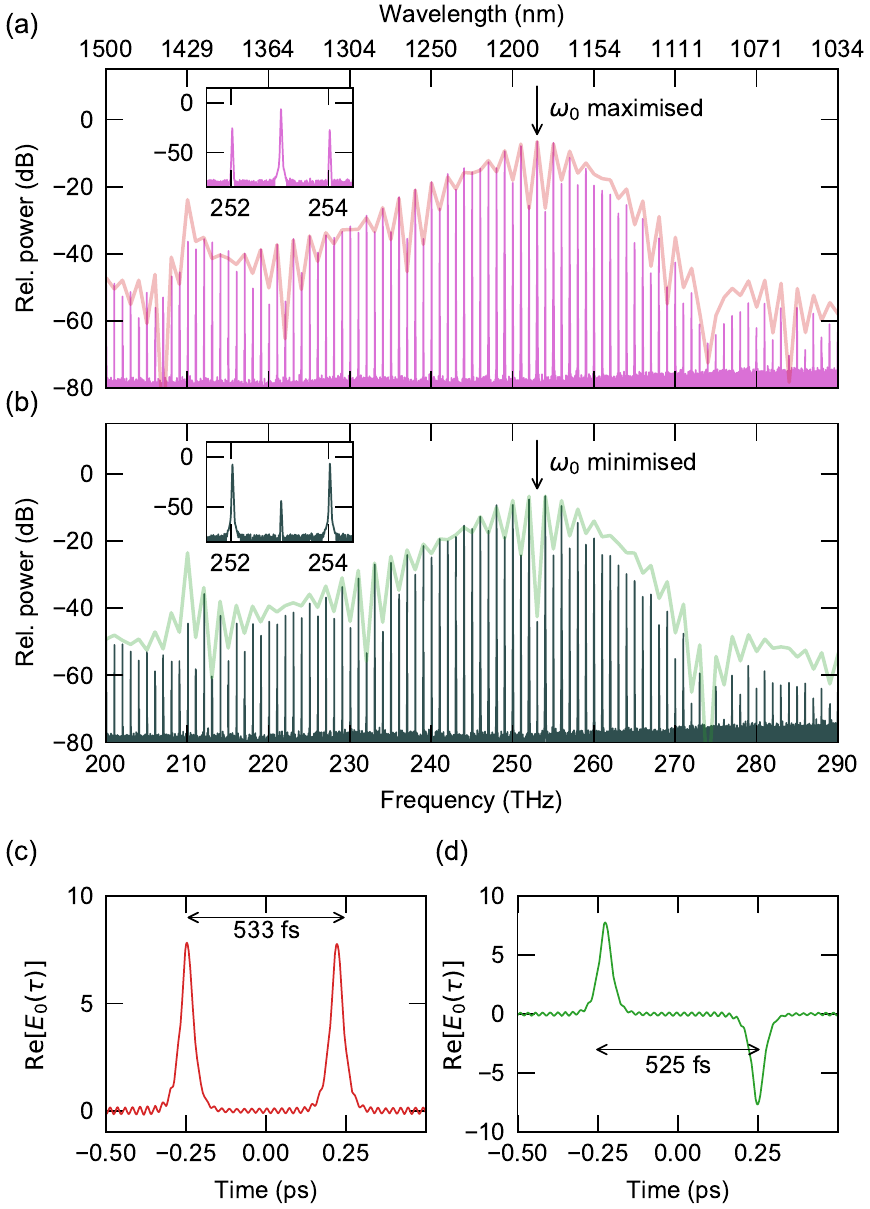}
	\caption{\textbf{Observations of multi-soliton interference.} (a, b) Comb spectra corresponding to coherent states with two PDCSs simultaneously circulating in the resonator. The red and green shaded curves in (a) and (b) depict out-coupled spectral envelopes of the two-soliton fields whose real part is shown in (c) and (d), respectively, created by linearly superposing two PDCS solutions with different relative delay and phase [Methods]. In (c), the solitons are in-phase and have a relative temporal separation of 533~fs, whilst in (d) the solitons are out-of-phase (relative phase $0.992 \pi$) and have a temporal separation of 525~fs. Insets in (a) and (b) show a zoomed-in view of the measured comb spectra around the degenerate FWM frequency 253~THz, highlighting how the degenerate FWM component at 253~THz is (a) maximised and (b) minimised.}
	\label{fig4}
\end{figure}

The emergence of the smooth comb state [Fig.~\ref{fig3}(e)] is associated with an abrupt drop in the photodetector signal recorded around 253~THz, giving rise to a noticeable step-like feature [Fig.~\ref{fig3}(f)]. Similar steps are well-known signatures of conventional CSs in monochromatically-driven Kerr resonators~\cite{herr_temporal_2014}. Moreover, as shown in Fig.~\ref{fig3}(e), the smooth spectral envelope observed in the step-region is in very good agreement with the spectrum of a 24~fs (full-width at half-maximum) PDCS derived from numerical modelling that use estimated experimental parameters [see Supplementary Figure~2]. The simulations faithfully reproduce the main features of the experimentally observed spectrum, including a strong dispersive wave peak at about 210~THz. We note that the prominent dip at about 275~THz arises due to the frequency-dependence of the pulley coupler~\cite{moille_broadband_2019},  which was taken into account \emph{ad hoc} when estimating the spectrum of the out-coupled PDCS [shown as blue curve in Fig.~\ref{fig3} -- see also Methods and Supplementary Figure~2].

It is interesting to note that, in addition to the frequency comb around the degenerate FWM frequency 253~THz, frequency combs arise also around both of the pump frequencies. These combs originate from FWM interactions between the pump fields and the comb lines around 253~THz, in a manner similar to spectral extension~\cite{zhang_spectral_2020,moille_ultra-broadband_2021, qureshi_soliton_2021} and two-dimensional frequency comb~\cite{MoillearXiv2023} schemes studied in the context of conventional Kerr CSs. The combs around the pump frequencies share the line spacing with the comb around 253~THz, but there is a constant offset between the pump and PDCS combs. In our experiments, this comb offset is directly observable in the optical spectrum [inset of Fig.~\ref{fig3}(e)] and found to be about 50~GHz $\pm$ 2~GHz (uncertainty defined by the optical spectrum analyzer resolution), which is in good agreement with the value of 49~GHz predicted by our modelling [see Methods]. All in all, given the considerable uncertainties in key experimental parameters (particularly dispersion and detunings), we find the level of agreement between the simulations and experiments remarkable.

The results shown in Fig.~\ref{fig3} are strongly indicative of PDCS generation in our experiments. Further confirmation is provided by observations of low-noise combs with complex spectral structures that afford a straightforward interpretation in terms of multi-PDCS states [Fig.~\ref{fig4}]. Specifically, whilst a single PDCS circulating in the resonator is expected to yield a smooth spectral envelope, the presence of two (or more) PDCSs results in a spectral interference pattern whose details depend upon the soliton's relative temporal delay and -- importantly -- phase.

Figures~\ref{fig4}(a) and (b) show selected examples of multi-soliton comb spectra measured in our experiments. Also shown as solid curves are spectral envelopes corresponding to fields with two linearly superposed, temporally delayed PDCSs [Figs.~\ref{fig4}(c) and (d) and Methods]. We draw particular attention to the fact that, in the measured data shown in Fig.~\ref{fig4}(b), the comb component at the degenerate FWM frequency $\omega_0$ is suppressed by about 40~dB compared to neighbouring lines, which is in stark contrast with results in Fig.~\ref{fig4}(a), in which the degenerate FWM component is dominant. This suppression is indicative of a relative phase shift of $0.992\pi$ between the two solitons [Fig.~\ref{fig4}(d)] -- a clear signature of PDCSs.

\section{Discussion}

We have shown theoretically, numerically, and experimentally that dispersive resonators with a purely Kerr-type $\chi^{(3)}$ nonlinearity can support parametrically driven cavity solitons under conditions of bichromatic driving. Our theoretical analysis has revealed the salient conditions that the system dispersion must meet to allow for PDCS persistence, with approximation-free numerical simulations confirming the fundamental viability of the scheme. Experimentally, we realise suitable dispersion conditions in a chip-integrated silicon nitride microresonator, observing low-noise frequency comb states that evidence PDCS generation. Significantly, our measurements show spectral interference patterns that indicate the co-existence of two localized structures with opposite phase -- a defining feature of PDCSs.

Our work fundamentally predicts and demonstrates that dispersive Kerr resonators can support a new type of dissipative structure -- the PDCS -- in addition to conventional Kerr CSs. We envisage that studying the rich nonlinear dynamics~\cite{leo_dynamics_2013, yu_breather_2017, lucas_breathing_2017}, interactions~\cite{jang_ultraweak_2013, cole_soliton_2017, wang_universal_2017}, and characteristics (including quantum~\cite{chembo_quantum_2016,bao_quantum_2021, yang2021squeezed, guidry_quantum_2022}) of pure-Kerr PDCSs will draw substantial future research interest, echoing the extensive exploration of conventional Kerr CS dynamics over the past decade~\cite{leo_dynamics_2013, yu_breather_2017, lucas_breathing_2017, jang_ultraweak_2013, cole_soliton_2017, wang_universal_2017, chembo_quantum_2016,bao_quantum_2021, yang2021squeezed, guidry_quantum_2022}.  In this context, to the best of our knowledge, the results reported in our work represent the first prediction and observation of dispersive wave emission by PDCSs in any physical system.

From a practical vantage, our scheme offers a route to generate PDCS frequency combs in foundry-ready, chip-integrated platforms with characteristics that are fundamentally different from those associated with conventional Kerr CSs. For example, forming between the two input frequencies, PDCSs could permit comb generation at spectral regions where direct pump lasers may not be available. Moreover, the lack of a dominating CW component at the PDCS carrier frequency alleviates the need for careful spectral shaping, and could result in fundamental advantages to noise characteristics. We emphasise that PDCSs are underpinned by phase-sensitive amplification~\cite{mecozzi_long-term_1994}, which can theoretically offer a sub-quantum-limited (squeezed) noise figure~\cite{McKinstrie_phase-sensitive_2004,slavik_all-optical_2010,tong_towards_2011,Zhang2008, Ye2021}. Finally, the fact that PDCSs come in two forms with opposite phase opens the doors to a new range of applications that require a binary degree of freedom, including all-optical random number generation and realisations of coherent optical Ising machines. Whilst the potential of PDCSs for such applications has been noted earlier~\cite{englebert_parametrically_2021}, our work provides for the first time a route for chip-integrated realisations with potential to CMOS-compatible mass manufacturing.

\section*{Methods}

\small

\subparagraph*{\hskip-10pt Simulation models.} We first describe the theoretical models that describe the dynamics of bichromatically driven Kerr resonators and that underpin simulation results in our work. Our starting point is a polychromatic Ikeda-like map, which we will use to derive an extended mean-field Lugiato-Lefever equation that has been used in previous studies~\cite{hansson_bichromatically_2014, taheri_optical_2017, zhang_spectral_2020,moille_ultra-broadband_2021, qureshi_soliton_2021, taheri_all-optical_2022}. To this end, we consider a Kerr resonator made out of a dispersive waveguide [with length $L$ and propagation constant $\beta(\omega)$] that is driven with two coherent fields with angular frequencies $\omega_\pm$ [see Fig.~\ref{fig1}(c)]. The evolution of the electric field envelope [referenced against the degenerate FWM frequency $\omega_0 = (\omega_++\omega_-)/2$] during the $m$th transit around the resonator is governed by the generalized nonlinear Schr\"odinger equation:
\begin{equation}
\frac{\partial E^{(m)}(z,\tau)}{\partial z} = i\hat{\beta}_\mathrm{S}\left(i\frac{\partial}{\partial\tau}\right)E^{(m)} + i\gamma |E^{(m)}|E^{(m)}. \label{Seq}
\end{equation}
Here $z$ is a coordinate along the waveguide that forms the resonator, $\tau$ is time in a reference frame that moves with the group-velocity of light at $\omega_0$, $\gamma$ is the Kerr nonlinearity coefficient and the dispersion operator
\begin{equation}
\hat{\beta}_\mathrm{S}\left(i\frac{\partial}{\partial\tau}\right) = \sum_{k\geq 2} \frac{\beta_k}{k!}\left(i\frac{\partial}{\partial\tau}\right)^k,
\label{dispop}
\end{equation}
with $\beta_k = d\beta/d\omega|_{\omega_0}$ the Taylor series expansion coefficients of $\beta(\omega)$ around $\omega_0$. Note that the single electric field envelope $E^{(m)}(\tau,z)$ contains all the frequency components pertinent to the nonlinear interactions, including the fields at the pump frequencies $\omega_\pm$ and the signal frequency at $\omega_0$. Note also that the Taylor series expansion coefficients $\beta_k$ are linked to the resonance frequency expansion coefficients in Eq.~\eqref{Dint} as $D_k \approx -D_1^{k+1} L \beta_k/(2\pi)$~\cite{pasquazi_micro-combs_2018}, such that
\begin{equation}
D_\mathrm{int}(\mu) \approx -\frac{D_1 L}{2\pi}\hat{\beta}_\mathrm{S}(\mu D_1).
\end{equation}

The Ikeda map consists of Eq.~\eqref{Seq} together with a boundary equation that describes the coupling of light into the resonator. Considering bichromatic driving, the boundary equation reads [see also Supplementary Note~1]:
\begin{align}
E^{(m+1)}(0,\tau) &= \sqrt{1-2\alpha} E^{(m)}(L,\tau)e^{-i\delta_0} \nonumber \\
 &+ \sqrt{\theta_+}E_\mathrm{in,+} e^{-i\Omega_\mathrm{p}\tau + imb_+} \nonumber \\
 &+ \sqrt{\theta_-}E_\mathrm{in,-} e^{i\Omega_\mathrm{p}\tau + imb_-}. \label{Boundary}
\end{align}
Here $\alpha$ is half of the fraction of power dissipated by the intracavity field over one round trip, $\delta_0 = 2\pi k - \beta(\omega_0) L$ is the linear phase detuning of the reference frequency $\omega_0$ from the closest cavity resonance (with order $k$), $E_\mathrm{in,\pm}$ are the complex amplitudes of the driving fields at $\omega_\pm$, respectively, $\Omega_\mathrm{p} = pD_1$ with $p$ a positive integer represents the angular frequency shifts of the pumps from the reference frequency $\omega_0$, and $\theta_\pm$ are the power transmission coefficients that describe the coupling of the driving fields into the resonator.  The coefficients $b_\pm$ allow us to introduce the phase detunings $\delta_\pm$ that describe the detunings of the pump frequencies from the cavity resonances closest to them (thus accounting for the fact that the frequency shift $\omega_0-\omega_\pm$ may not be an exact integer multiple of $D_1$):
\begin{equation}
b_\pm = \delta_\pm -\delta_0 +\hat{\beta}_\mathrm{S}(\pm\Omega_\mathrm{p})L. \label{bs}
\end{equation}
Note that the phase detunings $\delta$ described above are related to the frequency detunings of the corresponding carrier frequency $\omega$ from the closest cavity resonances at $\omega'$ as $\delta \approx 2\pi (\omega'-\omega)/D_1$.

Before proceeding, we note that, in our specific configuration, only two out of the three detuning terms introduced above ($\delta_0$ and $\delta_\pm$) are independent. This is because the degenerate FWM frequency is completely determined by the pump frequencies viz. $\omega_0 = (\omega_+ + \omega_-)/2$; therefore, the signal detuning $\delta_0$ can be written in terms of the pump detunings $\delta_\pm$ as [see Supplementary Note~2]:
\begin{equation}
\delta_0 = \frac{\delta_+ + \delta_- + L[\hat{\beta}_\mathrm{S}(\Omega_\mathrm{p})+\hat{\beta}_\mathrm{S}(-\Omega_\mathrm{p})]}{2}. \label{d0}
\end{equation}
Substituting this expression for $\delta_0$ into Eq.~\eqref{bs} yields $b_\pm = \pm b$, where
\begin{equation}
b = \frac{\delta_+ - \delta_- + L[\hat{\beta}_\mathrm{S}(\Omega_\mathrm{p})-\hat{\beta}_\mathrm{S}(-\Omega_\mathrm{p})]}{2}. \label{bcoef}
\end{equation}
It can be shown [see Supplementary Note~3] that this coefficient describes the offset, $\Delta f$, between the frequency combs forming around $\omega_0$ and $\omega_\pm$ viz.
\begin{equation}
\Delta f = \frac{|b|D_1}{(2\pi)^2}. \label{deltaf}
\end{equation}

\subparagraph*{\hskip-10pt PDCS theory.} All of the simulations presented in our work use the full Ikeda-like map defined by Eqs.~\eqref{Seq} and~\eqref{Boundary}. However, the system's ability to sustain PDCSs can be inferred more readily from the mean-field limit, obtained under the assumption that the intracavity envelope $E^{(m)}(z,\tau)$ evolves slowly over a single round trip (i.e., the cavity has a high finesse, and the linear and nonlinear phase shifts are all small). In this case, the Ikeda-like map described above can be averaged into the generalized Lugiato-Lefever mean-field equation similar to the one used, e.g., in refs.~\cite{zhang_spectral_2020,moille_ultra-broadband_2021, qureshi_soliton_2021, taheri_all-optical_2022}. We write the equation in normalized form as [see Supplementary Note~4]:
\begin{align}
\frac{\partial E(t,\tau)}{\partial t} &= \left[-1 + i(|E|^2 -\Delta_0)+ i\hat{\beta}\left(i\frac{\partial}{\partial\tau}\right)\right]E \label{LLN} \\
 &+ S_+ e^{-i\Omega_\mathrm{p}\tau + iat} \nonumber + S_- e^{i\Omega_\mathrm{p}\tau - iat}.  \nonumber
\end{align}
Here $t$ is a slow time variable that describes the evolution of the intracavity field over consecutive round trips (and is thus directly related to the index $m$ of the Ikeda-like map), \mbox{$S_\pm = E_\mathrm{in,\pm} \sqrt{\gamma L \theta_\pm/\alpha^3}$} are the normalized strengths of the driving fields, $\Delta_0 = \delta_0/\alpha$ is the normalized detuning of the signal field, and the normalized dispersion operator $\hat{\beta}$ is defined as Eq.~\eqref{dispop} but with normalized Taylor series coefficients $\beta_k\rightarrow d_k = [2\alpha/(|\beta_2|L)]^{k/2}\beta_kL/\alpha$. Finally, the coefficient
\begin{equation}
a = \frac{b}{\alpha} = \frac{\Delta_+ - \Delta_- + [\hat{\beta}(\Omega_\mathrm{p})-\hat{\beta}(-\Omega_\mathrm{p})]}{2}, \label{acoef}
\end{equation}
where $\Delta_\pm = \delta_\pm/\alpha$ are the normalized detunings of the external driving fields. To avoid notational clutter, we use the symbol $\Omega_\mathrm{p}$ to represent pump frequency shifts both in our dimensional and normalized equations.

We now make the assumption that the intracavity fields $E_\pm$ at the pump frequencies are homogeneous and stationary. (Note: this assumption is not used in any of our simulations.) To this end, we substitute the ansatz
\begin{align}
E(t,\tau) &= E_0(t,\tau) \label{anz}\\
&+ E_+e^{-i\Omega_\mathrm{p}\tau + ia t} + E_-e^{i\Omega_\mathrm{p}\tau - ia t}  \nonumber
\end{align}
into Eq.~\eqref{LLN}. We then assume further that the (soliton) spectrum around the degenerate FWM frequency (the Fourier transform of $E_0(t,\tau)$) does not exhibit significant overlap with the pump frequencies. This allows us to separate terms that oscillate with different frequencies, yielding the following equation for the signal field:
\begin{align}
\frac{\partial E_0(t,\tau)}{\partial t} &=\left[-1 + i(|E_0|^2 -\Delta_\mathrm{eff}) + i\hat{\beta}\left(i\frac{\partial}{\partial\tau}\right) \right]E_0 \nonumber \\
&+ 2iE_+E_- E_0^\ast, \label{PDNLSE}
\end{align}
where the effective detuning $\Delta_\mathrm{eff} = \Delta_0 - 2(Y_+ + Y_-)$ with $Y_\pm=|E_\pm|^2$ includes both linear and nonlinear (cross-phase modulation) phase shifts. Equation~\eqref{PDNLSE} has the precise form of the parametrically-driven nonlinear Schr\"odinger equation~\cite{bondila_topography_1995} with effective detuning $\Delta_\mathrm{eff}$ and parametric driving coefficient $\nu = 2iE_+E_-$.  Accordingly, assuming that the resonator group-velocity dispersion is anomalous at the signal frequency ($\beta_2 < 0$), the equation admits exact (parametrically-driven) soliton solutions of the form~\cite{englebert_parametrically_2021}:
\begin{equation}
E_0(\tau) = \sqrt{2}\zeta \text{sech}(\zeta\tau)e^{i(\phi + \theta)}, \label{PDsol}
\end{equation}
where $\cos(2\phi) = 1/|\nu|$, $\zeta = \sqrt{\Delta_\mathrm{eff}+|\nu|\sin(2\phi)}$, and \mbox{$\theta = \text{arg}[iE_+E_-]$}. It should be clear from the last term of Eq.~\eqref{PDNLSE} that \emph{all} of the frequency components of $E_0$ are parametrically driven. This is particularly evident when expanding the field as a Fourier series, $E_0(t,\tau)=\sum_n c_n(t) e^{-inD_1\tau}$: the equation of motion for each modal amplitude $c_n$ will include a parametric driving term $2iE_+E_-c_{-n}^\ast$.

Of course, the viability of sustaining the PDCS solution described by Eq.~\eqref{PDsol} in an actual bichromatically-driven Kerr resonator system is contingent on the applicability of the assumptions outlined above. As described in the main text, the assumption that the intracavity fields $E_\pm$ at the pump frequencies are homogeneous and stationary lead to the requirements of dispersive walk-off and suppression of modulation instabilities. The requirement for phase-matching of the degenerate FWM process ensues from the fact that stable PDCS solutions generically exist only if the effective detuning $\Delta_\mathrm{eff}$ is sufficiently small~\cite{bondila_topography_1995}. Indeed, recalling Eq.~\eqref{d0}, we have
\begin{equation}
\Delta_\mathrm{eff} = \frac{\Delta_+ + \Delta_- + \hat{\beta}(\Omega_\mathrm{p}) + \hat{\beta}(-\Omega_\mathrm{p})}{2} - 2\left(Y_+ + Y_-\right). \label{Deltaeeff}
\end{equation}
Considering typical parameters, $\Delta_\mathrm{eff}$ and $|\nu| = 2\sqrt{Y_+Y_-}$ are of the order of unity for stable solitons to exist~\cite{englebert_parametrically_2021, bondila_topography_1995}, while the detunings $\Delta_\pm$ can be assumed small to ensure that sufficient intracavity powers $Y_\pm$ can be attained without excessive driving powers $X_\pm = |S_\pm|^2$. This implies, then, that the pump frequency shift $\Omega_\mathrm{p}$ must satisfy $[\hat{\beta}(\Omega_\mathrm{p}) + \hat{\beta}(-\Omega_\mathrm{p})]\approx 0$. Unpeeling the normalization, and converting to the integrated dispersion defined as Eq.~\eqref{Dint} of the main text, shows that this condition is equivalent with the linear phase-matching of degenerate FWM: $D_\mathrm{int}(p) + D_\mathrm{int}(-p)\approx 0$.

\subparagraph*{\hskip-10pt Resonator used in experiments.} The chip-integrated microring resonator used in our experiments was fabricated in a commercially-available foundry service. The resonators are made of a 690~nm-thick layer of silicon nitride that is fully embedded in fused silica. The ring has a width of 850~nm and a radius of 23~$\mu$m, thus yielding a round trip length $L = 144.5~\mathrm{\mu m}$. Light is coupled into the ring via a 460~nm-wide integrated bus waveguide, with a 32~$\mu$m-long pulley-coupler ensuring good coupling at all the different frequencies of interest ($\omega_0$, $\omega_\pm$). The resonator has intrinsic and loaded $Q$-factors of $1.5\times10^6$ and $0.75\times10^6$, respectively, corresponding to a finesse of $\mathcal{F} \approx 3000$ and a resonance linewidth of \mbox{$\Delta f_\mathrm{r} \approx 300~\mathrm{MHz}$}. The chip has an input-to-output insertion loss of about 5.6~dB at 980~nm and 8.4~dB at 1550~nm.

\subparagraph*{\hskip-10pt Resonator dispersion and thermal nonlinearity.} The theoretically estimated resonator dispersion [orange curve shown in Fig.~\ref{fig3}(b)] was obtained in two steps. We first calculated the theoretical resonance frequencies using finite-element modelling, and then slightly modified that data [see Supplementary Figure~3 for a comparison of the two integrated dispersion curves] to match the PDCS simulations to experimentally obtained spectra. Experimentally, we characterized the dispersion at various spectral regions by measuring the resonance frequencies using a set of widely tunable lasers and a high-resolution wavemeter. Unfortunately, the unavailability of a suitable laser around the degenerate FWM frequency (253~THz) prevented us from directly probing the dispersion at that frequency.

Because we are not able to probe the dispersion around 253~THz, it is not possible to unequivocally compare experimentally measured dispersion with our theoretical estimate. This is because the integrated dispersion $D_\mathrm{int}$ depends upon the precise resonance frequency $\omega'_0$ and the free-spectral range [$D_1/(2\pi)$] at $\omega'_0$, which we are unable to probe experimentally. To nonetheless show that our measurements at different spectral regions are \emph{consistent} with our theoretical estimate, we can use nonlinear least-squares to fit our experimental data to the theoretical data, and in doing so obtain experimental estimates for $\omega'_0$ and $D_1$, which then allows us to compute the integrated dispersion. The blue dots in Fig.~\ref{fig3}(a) were obtained using this procedure. The fitting also provides the one-standard-deviation errors for the parameter estimates, $\Delta\omega'_0$ and $\Delta D_1$, which then allows us to compute the fitting errors for $\Delta D_\mathrm{int}(\mu)$ and $\Delta\delta\omega(\mu)$. We find that the maximum (across relative mode order $\mu$) error in the estimated $D_\mathrm{int}$ is $\text{max}[\Delta D_\mathrm{int}(\mu)/(2\pi)] \approx 0.50~\mathrm{GHz}$, yielding $\text{max}[\Delta \delta\omega(\mu)/(2\pi)] \approx 0.35~\mathrm{GHz}$. These errors are smaller than the markers used in Figs.~\ref{fig3}(b) and (c), which is why errorbars are not shown.

Due to the resonator's small size, it exhibits a strong thermal nonlinearity~\cite{carmon_dynamical_2004}. We leverage this effect to achieve self-stabilization, such that the input lasers can remain free-running but still maintain near-constant detunings. In addition, the thermal nonlinearity causes the resonance frequencies to shift over several GHz as the pump laser(s) are tuned into resonance [see e.g. Fig.~\ref{fig3}(f)], which we suspect is key to achieving phase-matched operation (and thus PDCS generation). We also note that the thermal nonlinearity may influence the resonator dispersion directly~\cite{moille_integrated_2022}; whilst this effect is generally weak (and under-examined), it is possible that it also influences the precise phase-matching conditions, thus playing a role in our experiments. A detailed study on the impact of the thermal nonlinearity to PDCS generation is beyond the scope of our present work.

\subparagraph*{\hskip-10pt Simulation parameters.} The simulations in Fig.~2 assume a critically-coupled ($\alpha = \theta$) resonator with a round trip length $L \approx 8.3~\mathrm{mm}$, nonlinearity coefficient $\gamma = 1.2~\mathrm{W^{-1}km^{-1}}$, and finesse $\mathcal{F} = \pi/\alpha = 5000$. The driving fields are positioned at an angular frequency shift $\pm\Omega_\mathrm{p} = 2\pi\times 30.4~\mathrm{THz}$ with respect to the degenerate FWM frequency, corresponding to relative mode number $p = 1217$. The dispersion coefficients are $\beta_2 = -5~\mathrm{ps^2/km}$, $\beta_3 = 0.45
\mathrm{ps^3/km} $ and $\beta_4 = 1.6\times 10^{-3}\mathrm{ps^4/km}$, corresponding to $D_2/(2\pi) = 4.06~\mathrm{kHz}$, $D_3/(2\pi) = -57.90~\mathrm{Hz}$ and $D_4 = -0.03~\mathrm{Hz}$.

The above parameters yield an effective (normalised) driving strength $|\nu| = 1.37$ and detuning $\Delta_\mathrm{eff} = 1.2$ which are known to be in the regime of soliton existence~\cite{englebert_parametrically_2021}. As a matter of fact, the above parameters were found by looking for the driving powers and frequency shifts that yield these particular values for the driving strength and detuning.

The simulations in Fig.~\ref{fig3} and~\ref{fig4} use experimental values quoted in the main text or in the resonator description above, with the addition that the nonlinearity coefficient was set to $\gamma = 1~\mathrm{W^{-1}m^{-1}}$. The pump detunings were chosen such that, in Fig.~\ref{fig3}, the effective driving strength $|\nu| = 1.28$ and $\Delta_\mathrm{eff} = 6$, and in Fig.~\ref{fig4}, $|\nu| = 1.15$ and $\Delta_\mathrm{eff} = 5$. The effective detunings were coarsely tuned so as to match the simulations to the experimentally measured spectra. The simulation outcomes are not sensitive to the particular values of the driving strength $\nu$ used.

With the parameters used to obtain the simulation results in Fig.~\ref{fig3}, the coefficient $b$ defined in Eq.~\eqref{bcoef} was $b=-0.307$, yielding a comb frequency offset of $\Delta f = 49~\mathrm{GHz}$ from Eq.~\eqref{deltaf}.

\subparagraph*{\hskip-10pt Frequency-dependent coupling.} All the simulations reported in our manuscript have been obtained using the model defined by Eqs.~\eqref{Seq} and~\eqref{Boundary}. However, as explained in the main text [see also Supplementary Figure~2], when comparing against experimentally measured spectra [Figs.~\ref{fig3} and~\ref{fig4}], the simulation outputs were post-processed to account for the frequency-dependent coupling, thus providing an estimate for the out-coupled spectrum. This was achieved by multiplying the simulated intracavity spectra with the frequency-dependent coupling coefficient [Supplementary Figure~2] obtained from rigorous coupled-mode simulations~\cite{moille_broadband_2019}. These coupled-mode simulations assumed the coupler length to be 31.25~$\mu$m, which was found to provide a better agreement with our experiments compared to the design value of $32$~$\mu$m. This discrepancy is reasonable in terms of fabrication tolerances given the high sensitivity to the phase mismatch between the ring and waveguide modes and that any small discrepancy in the side-wall angle or waveguide width could cause a smaller effective pulley. However we note that the obtained length is well within fabrication tolerance of deep-UV stepper fabrication. Note that the frequency-dependent coupling was not included explicitly in our numerical simulation model for the sake of simplicity.

\subparagraph*{\hskip-10pt Multi-soliton states.} Because of pump depletion and finite dispersive walk-off, the PDCSs carve a depletion region onto the intracavity fields at the pump frequencies [see Fig.~\ref{fig1}(f)]. These depletion regions are the time-domain manifestations of the frequency combs that form around the pump frequencies, and they give rise to long-range soliton interactions. Compounded by the system's periodic boundary conditions, stable multi-soliton states only exist at selected relative delays (or not at all) in our simulations. On the other hand, it is well-known (from studies of conventional Kerr CSs) that experimental systems exhibit imperfections (e.g. avoided mode crossings) which, along with oscillatory tails from dispersive waves, force multi-soliton states to only manifest themselves at some prescribed relative delays~\cite{cole_soliton_2017, wang_universal_2017}. Because the PDCSs in our simulations exhibit long-range coupling, it is not possible to obtain a simulation of a multi-soliton state with the same relative delays as in our experiments, unless one has access to full details of the experimental system (including dispersion that captures possible avoided mode crossings), which we do not have.

Because of the above, the theoretical PDCS fields in Fig.~\ref{fig4}(c) and~(d) were created from a single steady-state PDCS -- obtained via simulations of Eqs.~\eqref{Seq} and~\eqref{Boundary}. Specifically, the two-soliton fields were obtained by linearly adding together two replicas of the single steady-state PDCS state, with the relative delay $(\Delta\tau)$ and phase $(\Delta\phi)$ between the replicas inferred from nonlinear least squares fitting to the experimentally observed spectral interference pattern. For both in- and out-of-phase states, our fitting algorithm yields two possible configurations $(\Delta \tau, \Delta\phi)$ that identically minimise the sum of the squared residuals. For the in-phase configuration, these are $(533~\mathrm{fs},1\times10^{-3}\pi)$ and $(467~\mathrm{fs},3\times10^{-4}\pi)$,  and for the out-of-phase configuration we have $(525~\mathrm{fs}, 0.99\pi)$ and $(475~\mathrm{fs}, 1.01\pi)$. In Figs.~\ref{fig4}(c) and (d), we plot the configurations associated with the larger delay. The one-standard-deviation errors for the fits are all smaller than $(0.4~\mathrm{fs}, 0.01\pi)$.

\section*{Acknowledgements}
\noindent G.~M. and K.~S. acknowledge support from the NIST-on-a-chip program. J.~F. acknowledges the CNRS (IRP WALL-IN project).

\section*{Author Contributions}

\noindent G.~M. performed all the experiments and assisted in the interpretation of the results. M.~L. and D.~P contributed to the theoretical development of the scheme and performed initial simulations to confirm the fundamental viability of the scheme. N.~E. and F.~L. provided guidance on parametrically-driven soliton theory. J.~F. assisted in the interpretation of Kerr cavity physics. K.~S. supervised and obtained funding for the experiments. M.~E. developed the theory, performed the simulations, and wrote the manuscript with input from all the authors.
\section*{Data availability}

\noindent The data that support the plots within this paper and other findings of this study are available from M.E. upon reasonable request.

\section*{Competing financial interests}

\noindent The authors declare no competing financial interests.

\bibliographystyle{bibstyle2nonotes}

\clearpage

\section{Supplementary Note~1: Derivation of the Ikeda map}
We present here a heuristic derivation of the Ikeda-like map used in our numerical simulations [Eq.~(5) of the main manuscript]. Including the rapid temporal oscillations at $\omega_0 = (\omega_+ + \omega_-)/2$, where $\omega_\pm$ are the input frequencies, the intracavity electric field during the $m^\mathrm{th}$ cavity transit is written as $E^{(m)}(z,\tau)\exp[-i\omega_0 T]$, where $\tau$ is time in a co-moving reference frame defined as $\tau = T-z/v_\mathrm{g}$ with $T$ absolute laboratory time, $z$ the coordinate along the waveguide that forms the resonator, $v_\mathrm{g}$ the group velocity of light at $\omega_0$, and $E^{(m)}(z,\tau)$ is the slowly-varying electric field envelope that follows the generalized nonlinear Schr\"odinger equation [Eq.~(2) of the main manuscipt]. The boundary equation for the full electric field can then be written as
\begin{align}
E^{(m+1)}(0,\tau)e^{-i\omega_0 T} &= \sqrt{1-2\alpha} E^{(m)}(L,\tau)e^{-i\delta_0-i\omega_0 T} \nonumber \\
 &+ \sqrt{\theta_+}E_\mathrm{in,+} e^{-i\omega_+T} \nonumber \\
 &+ \sqrt{\theta_-}E_\mathrm{in,-} e^{-i\omega_-T}, \label{Boundary2}
\end{align}
where $\delta_0 = 2\pi k - \beta(\omega_0) L$ is the linear phase detuning of the reference frequency $\omega_0$ from the closest cavity resonance (with order $k$), and $\omega_\pm$ are the frequencies of the pump fields. Multiplying each side with $\exp[i\omega_0 T]$ and replacing $T = \tau + mL/v_\mathrm{g}= \tau + mt_\mathrm{R}$ where $t_\mathrm{R}$ is the round trip time, yields
\begin{align}
E^{(m+1)}(0,\tau) &= \sqrt{1-2\alpha} E^{(m)}(L,\tau)e^{-i\delta_0} \nonumber \\
 &+ \sqrt{\theta_+}E_\mathrm{in,+} e^{-i\Omega_\mathrm{p}\tau + i(\omega_0-\omega_+)mt_\mathrm{R}} \nonumber \\
 &+ \sqrt{\theta_-}E_\mathrm{in,-} e^{i\Omega_\mathrm{p}\tau + i(\omega_0-\omega_-)mt_\mathrm{R}}. \label{Boundary3}
\end{align}
Note that, in the above formulation, the co-moving time variable $\tau$ should be understood as the ``fast time'' that describes the envelope of the intracavity electric field over a single round trip, i.e., the distribution of the envelope within the resonator~\cite{pasquazi_micro-combs_2018}. As such, the $\tau$ variable spans a single round trip time of the resonator, and the intracavity envelope must obey periodic boundaries within that range. These conditions stipulate that the frequency variable $\Omega_\mathrm{p} = 2\pi p \times \text{FSR}$, where $p$ is a positive integer and $\text{FSR} = t_\mathrm{R}^{-1}$ is the free-spectral range of the resonator. The fact that the frequency difference $(\omega_0-\omega_\pm)/(2\pi)$ may not, in general, be an integer multiple of the FSR is captured by the additional phase shifts accumulated by the driving fields with respect to the intracavity field from round trip to round trip.

To link the frequency differences $\omega_0-\omega_\pm$ to the respective phase detunings, we first recall that the phase detuning of a driving field with frequency $\omega$ from a cavity resonance at $\omega'$ obeys $\delta\approx (\omega'-\omega)t_\mathrm{R}$. We can thus write
\begin{equation}
\omega_0-\omega_\pm \approx  \omega'_0-\frac{\delta_0}{t_\mathrm{R}} -  \omega'_{\pm} + \frac{\delta_\pm}{t_\mathrm{R}}, \label{freqdif}
\end{equation}
where the frequency variables with (without) apostrophes refer to resonance (pump) frequencies. We next write the resonance frequencies as $\omega'_q = \omega'_0 + q D_1 + \hat{D}_\mathrm{int}(q)$, where $D_1 = 2\pi\text{FSR}$ with $\text{FSR} = t^{-1}_\mathrm{R}$ the free-spectral range of the cavity (at $\omega'_0$), $q$ an integer that represents the mode index (with $\omega_0$ corresponding to $q = 0$), and the integrated dispersion
\begin{equation}
D_\mathrm{int}(q) = \sum_{k\geq 2} \frac{D_k}{k!}q^k,
\label{DintS}
\end{equation}
where $D_k$ are the expansion coefficients. Assuming that the resonance frequencies $\omega'_\pm$ are associated with indices $\pm p$ (with $p>0$), respectively, we can write Eq.~\eqref{freqdif} as
\begin{equation}
\omega_0-\omega_\pm \approx  -\frac{\delta_0}{t_\mathrm{R}} \mp p D_1 - D_\mathrm{int}(\pm p) + \frac{\delta_\pm}{t_\mathrm{R}}. \label{freqdif2}
\end{equation}
The second term on the right-hand-side of Eq.~\eqref{freqdif2} can be ignored, as it yields an integer multiple of $2\pi$ when used in Eq.~\eqref{Boundary3}. Next, we use the fact~\cite{pasquazi_micro-combs_2018} that the coefficients $D_k$ with $k\geq 2$ can be linked to the Taylor series expansion coefficients of the propagation constant $\beta(\omega)$ viz. $D_k \approx - D_1^kL\beta_k/t_\mathrm{R}$. This allows us to write the integrated dispersion corresponding to the resonance frequencies $\omega'_\pm$ as
\begin{align}
D_\mathrm{int}(\pm p) &\approx \sum_{k\geq 2} -\frac{D_1^kL\beta_k}{t_\mathrm{R}k!}(\pm p)^k, \\
& = -\frac{L}{t_\mathrm{R}}\sum_{k\geq 2} \frac{\beta_k}{k!}(\pm D_1 p)^k.
\label{Dint2}
\end{align}
Using Eq.~\eqref{Dint2} in Eq.~\eqref{freqdif2} and substituting the latter into Eq.~\eqref{Boundary3} yields the Ikeda-like map described by Eq.~(5) of the main manuscript with coefficients $b_\pm$ as defined in Eq.~(6) of the manuscript, and the pump frequency shift $\Omega_\mathrm{p} = D_1 p = 2\pi p \times \text{FSR}$. We also note that the map can be straightforwardly extended to include arbitrarily many driving fields following the procedure above.

\section{Suplementary Note~2: signal detuning}
\label{apA2}
To derive the relationship between the parametric signal detuning $\delta_0$ and the pump detunings $\delta_\pm$ (i.e., Eq.~(7) of the main manuscript), we write out the parametric signal frequency as
\begin{equation}
\omega_0 = \frac{\omega_+ + \omega_-}{2} = \frac{\omega_+' - \Delta\omega_+ + \omega_-'- \Delta\omega_-}{2},
\end{equation}
where $\omega_\pm'$ are the resonance frequencies closest to the pump frequencies and $\Delta\omega_\pm$ are the angular frequency detuning of the pump frequencies from those resonances. Substituting $\Delta\omega_\pm = \delta_\pm/t_\mathrm{R}$ and expanding the pump resonance frequencies as $\omega_\pm' = \omega_0' \pm pD_1 + \hat{D}_\mathrm{int}(\pm p)$ yields
\begin{equation}
\omega_0 = \omega_0' - \frac{\delta_+ + \delta_-}{2t_\mathrm{R}} + \frac{\hat{D}_\mathrm{int}(p) + \hat{D}_\mathrm{int}(-p)}{2}.
\end{equation}
Then using Eq.~\eqref{Dint2} and rearranging, we obtain
\begin{align}
\delta_0 &= (\omega_0'-\omega_0)t_\mathrm{R} \nonumber\\
&= \frac{\delta_+ + \delta_- +  L[\hat{D}_\mathrm{S}(\Omega_\mathrm{p})+\hat{D}_\mathrm{S}(-\Omega_\mathrm{p})]}{2}.
\end{align}
This is Eq.~(7) of the main manuscript.

\section{Suplementary Note~3: comb offset}
The boundary equation of the Ikeda-like map [Eq.~(5) of the main manuscript] shows that the driving fields experience an additional relative phase shift per round trip determined by the coefficient
\begin{equation}
b = \frac{\delta_+ - \delta_- + L[\hat{\beta}_\mathrm{S}(\Omega_\mathrm{p})-\hat{\beta}_\mathrm{S}(-\Omega_\mathrm{p})]}{2}. \label{bcoefS}
\vspace{2pt}
\end{equation}
Using $\delta_\pm = (\omega'_\pm - \omega_\pm)t_\mathrm{R}$ and $L\hat{\beta}_\mathrm{S}(\pm\Omega_\mathrm{p})=-t_\mathrm{R} D_\mathrm{int}(\pm p)$, we obtain
\begin{align}
b = t_\mathrm{R}\left[ \frac{\omega'_+ - \omega'_-}{2} - \frac{\omega_+ - \omega_-}{2} - \frac{D_\mathrm{int}(p) -D_\mathrm{int}(-p)}{2} \right].
\end{align}
By then using $D_\mathrm{int}(p) - D_\mathrm{int}(-p) = \omega'_+ - \omega'_- - 2pD_1$, we obtain
\begin{align}
b = t_\mathrm{R}\left[pD_1 - \frac{\omega_+-\omega_-}{2}\right].
\end{align}

Given that $\omega_0 = (\omega_+ + \omega_-)/2$, we recognise $(\omega_+-\omega_-)/2$ as the angular frequency shift between the pump at $\omega_+$ and the degenerate FWM signal at $\omega_0$. Moreover, because the integer $p$ corresponds to mode number of the driven mode $\omega'_+$ relative to $\omega'_0$, we may write
\begin{equation}
b = t_\mathrm{R}\Delta\omega,
\end{equation}
where $\Delta\omega$ is the angular frequency difference between the pump at $\omega_+$ and the closest component of the frequency comb (with spacing $D_1/(2\pi)$) that forms around $\omega_0$. Defining the ordinary comb offset as $\Delta f = |\Delta\omega/(2\pi)|$, and using the fact that the combs that form around $\omega_0$ and $\omega_\pm$ have the same spacing, we obtain
\begin{equation}
\Delta f = \frac{|b|}{2\pi t_\mathrm{R}} = \frac{|b|D_1}{(2\pi)^2},
\end{equation}
where we used $D_1 = 2\pi/t_\mathrm{R}$.

\section{Supplementary Note~4: Lugiato-Lefever equation and normalization}
\label{apB}
Under the assumption that the intracavity envelope $E^{(m)}(z,\tau)$ evolves slowly over a single round trip (i.e., the cavity has a high finesse, and the linear and nonlinear phase shifts are all small), the Ikeda-like map described by Eqs.~(2) and~(5) of the main manuscript can be averaged into a single mean-field equation. The derivation is well-known~\cite{haelterman_dissipative_1992}, proceeding by integrating Eq.~(2) using a single step of the forward Euler method to obtain $E^{(m)}(L,\tau)$, which is then substituted into Eq.~(5). After linearizing with respect to $\delta_0$ and $\alpha$ and introducing the \emph{slow} time variable $t = mt_\mathrm{R}$ (such that the round trip index $m = t/t_\mathrm{R}$), one obtains:
\begin{align}
t_\mathrm{R}\frac{\partial E(t,\tau)}{\partial t} &= \left[-\alpha + i(\gamma L |E|^2 -\delta_0)+ iL\hat{\beta}_\mathrm{S}\left(i\frac{\partial}{\partial\tau}\right)\right]E \nonumber \\
 &+ \sqrt{\theta_+}E_\mathrm{in,+} e^{-i\Omega_\mathrm{p}\tau + ib_+t/t_\mathrm{R}} \nonumber \\
 &+ \sqrt{\theta_-}E_\mathrm{in,-} e^{i\Omega_\mathrm{p}\tau + ib_-t/t_\mathrm{R}}. \label{LLE}
\end{align}
To obtain the normalized Eq.~(10) of the main manuscript, we first introduce the variable transformations $\tau\rightarrow \tau \sqrt{2\alpha/(|\beta_2|L)}$, $t\rightarrow \alpha t/t_\mathrm{R}$, $\Omega_\mathrm{p}\rightarrow \Omega_\mathrm{p}\sqrt{|\beta_2|L/(2\alpha)}$ and $E\rightarrow E \sqrt{\gamma L/\alpha}$, yielding
\begin{align}
\frac{\partial E(t,\tau)}{\partial t} &= \left[-1 + i(|E|^2 -\Delta_0)+ i\hat{\beta}\left(i\frac{\partial}{\partial\tau}\right)\right]E \label{LLN2} \\
 &+ S_+ e^{-i\Omega_\mathrm{p}\tau + ia_+t} \nonumber + S_- e^{i\Omega_\mathrm{p}\tau + ia_-t},  \nonumber
\end{align}
where $S_\pm = E_\mathrm{in,\pm} \sqrt{\gamma L \theta_\pm/\alpha^3}$, and $\Delta_0 = \delta_0/\alpha$. The normalized dispersion operator $\hat{\beta}$ is defined as
\begin{equation}
\hat{\beta}\left(i\frac{\partial}{\partial\tau}\right) = \sum_{k\geq 2} \frac{d_k}{k!}\left(i\frac{\partial}{\partial\tau}\right)^k,
\label{dispop2}
\end{equation}
where the normalized dispersion coefficients are given by
\begin{equation}
d_k = \frac{\beta_k L}{\alpha}\left(\frac{2\alpha}{|\beta_2| L}\right)^{k/2}.
\end{equation}
Finally, the coefficients $a_\pm = \Delta_\pm - \Delta_0 + \hat{\beta}(\pm \Omega_\mathrm{p})$, where $\Delta_\pm = \delta_\pm/\alpha$ are the normalized detunings of the external driving fields. Note that, for the particular configuration considered in our work, where the signal frequency $\omega_0$ is strictly linked to the pump frequencies $\omega_\pm$ via $\omega_0 = (\omega_+ + \omega_-)/2$, the coefficients $a_\pm = \pm a$, where $a$ is defined by Eq.~(11) of the main manuscript.

\clearpage

\begin{figure*}[!t]
 \centering
\includegraphics[width = 0.9\textwidth, clip=true]{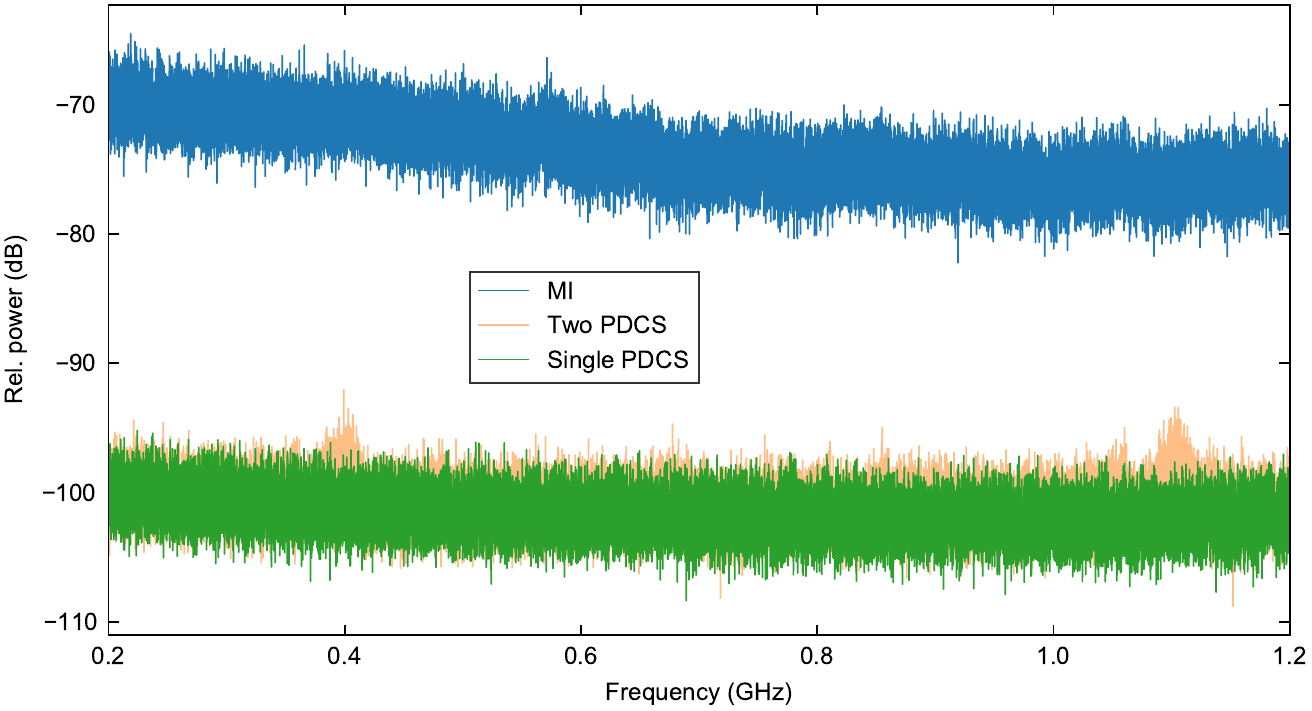}
 \caption{\textbf{Supplementary Figure 1}. Low-frequency photodetector noise measured on an electronic spectrum analyzer when the system is operating in the MI (blue curve), single PDCS (green curve), or two-PDCS (orange curve) regimes.}
 \label{figS2}
\end{figure*}

\begin{figure*}[!t]
 \centering
\includegraphics[width = 0.9\textwidth, clip=true]{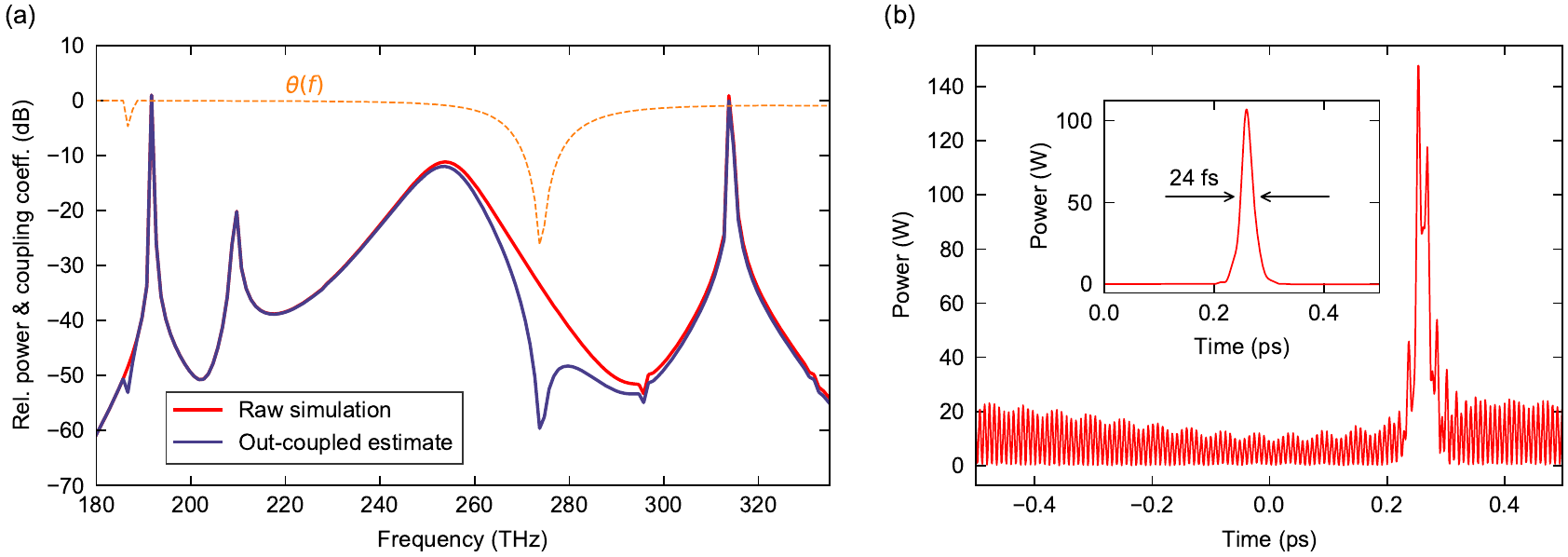}
 \caption{\textbf{Supplementary Figure 2}. (a) Red solid curve shows the optical spectrum obtained directly from our numerical simulations that assume a frequency-independent coupling coefficient $\theta_0$. The orange dashed curve shows the frequency-dependence of the coupling coefficient $\theta(f) = \theta_\mathrm{full}(f)/\theta_0$ estimated for our experiments, assuming a pulley length of 31.25~$\mu$m. The blue solid curve shows the estimated out-coupled spectrum, obtained by multiplying the simulated spectrum with $\theta(f)$. The data corresponding to the blue curve is identical to the curve shown in Fig.~3(e) of our manuscript. (b) Temporal intensity profile of the steady-state PDCS corresponding to our simulations. Inset shows the corresponding intensity profile after the pump fields have been filtered out.}
 \label{figS1}
\end{figure*}

\begin{figure*}[!t]
 \centering
\includegraphics[width = 0.9\textwidth, clip=true]{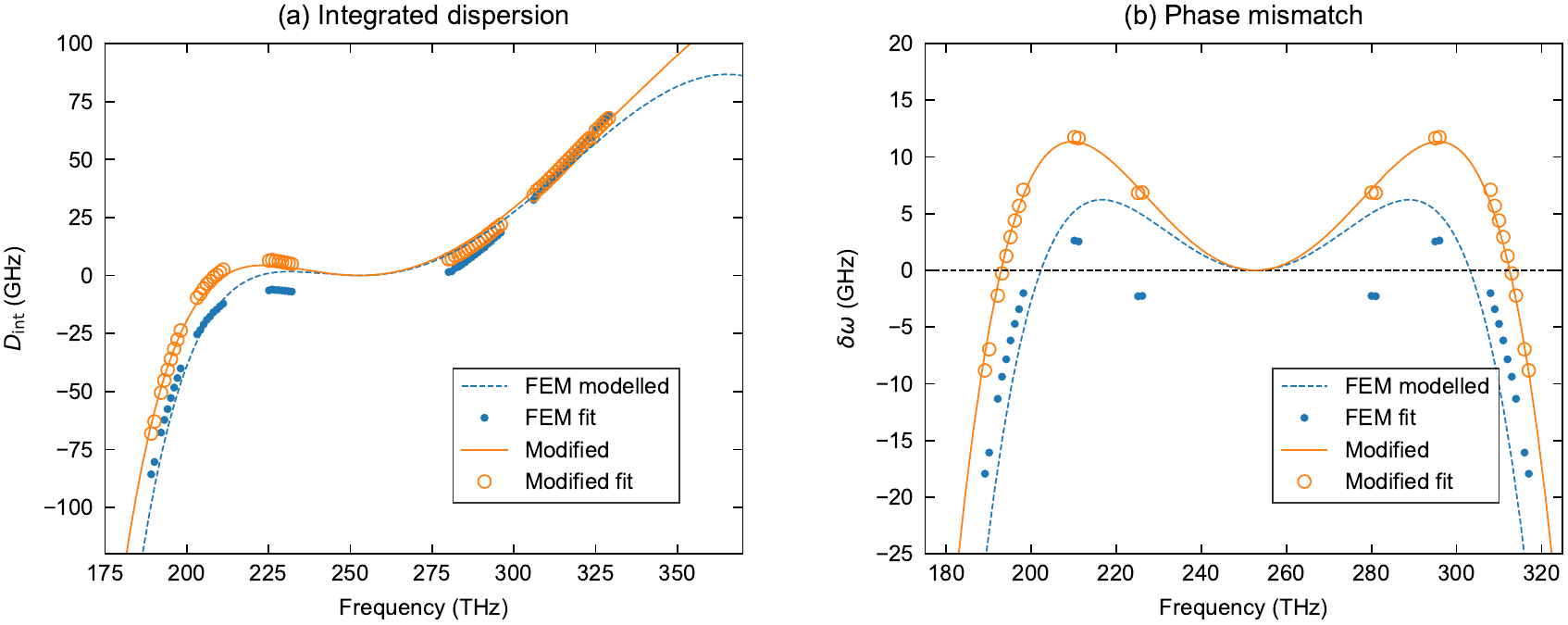}
 \caption{\textbf{Supplementary Figure 3}. (a) Blue dashed curve shows the integrated dispersion at 253~THz obtained from finite-element modelling (FEM) data, whilst blue dots show data extrapolated from experimental resonance measurements when fitted on the FEM data. Solid orange curve shows the modified integrated dispersion used to obtain the simulation results shown in Fig.~3 and~4 of our main manuscript. Open orange circles show data extrapolated from experimental resonance measurements when fitted on the modified dispersion data. (b) Phase-mismatch of the degenerate four-wave-mixing process computed from the different data presented in (a). Note that the orange curves and open circles portray the same data as shown in Fig.~3(b) and (c) of the main manuscript.}
 \label{figS3}
\end{figure*}

\end{document}